\documentclass[]{jfm}

\usepackage{graphicx}
\usepackage{newtxtext}
\usepackage{newtxmath}
\usepackage{natbib}
\usepackage{hyperref}
\usepackage{color}
\hypersetup{
    colorlinks = true,
    urlcolor   = blue,
    citecolor  = black,
}

\newcommand{\RomanNumeralCaps}[1]

\def\bk{\boldsymbol{k}}
\def\e{\mathrm{e}}
\def\i{\mathrm{i}}
\def\d{\mathrm{d}}
\defcitealias{kafiabad_savva_vanneste_2019}{KSV}

\renewcommand\epsilon{\varepsilon}

\usepackage{xcolor ,soul}

\title{Inertia-gravity-wave diffusion by geostrophic turbulence: the impact of flow time dependence}

\author{Michael R. Cox\aff{1}
  \corresp{\email{michael.cox@ed.ac.uk}},
  Hossein A. Kafiabad\aff{1}
 \and Jacques Vanneste\aff{1}}

\affiliation{\aff{1}School of Mathematics and Maxwell Institute for Mathematical Sciences, University of Edinburgh,
Edinburgh EH9 3FD, UK}

\begin{document}
\maketitle

\begin{abstract}
The scattering of three-dimensional inertia-gravity waves by a turbulent geostrophic flow  leads to the redistribution of their action  through what is approximately a diffusion process in wavevector space. The corresponding diffusivity tensor was obtained by \citeauthor*{kafiabad_savva_vanneste_2019} (2019, \textit{J. Fluid Mech.}, \textbf{869}, R7) under the assumption of a time-independent geostrophic flow. We relax this assumption to examine how the weak diffusion of wave action across constant-frequency cones that results from the slow time dependence of the geostrophic flow affects the   distribution of wave energy. We find that the stationary wave-energy spectrum that arises from a single-frequency wave forcing is localised within a thin boundary layer around the constant-frequency cone, with a thickness controlled by the acceleration spectrum of the geostrophic flow. We obtain an explicit analytic formula for the wave-energy spectrum which shows good agreement with the results of a high-resolution simulation of the Boussinesq equations.

\end{abstract}



\section{Introduction}

Atmospheric and oceanic inertia-gravity waves (IGWs) propagate in a complex turbulent flow which is in approximately geostrophic and hydrostatic balance. The inhomogeneities of this flow result in the scattering of IGWs which  redistributes their energy across wavevector space. This process has long been thought to play a role in the energetics of the atmosphere and ocean and it has been modelled using a range of approximations \citep*[see][]{mtfller_1976, MULLER197749, InteractionbetweenInternalWavesandMesoscaleFlow,mull-et-al, savva_kafiabad_vanneste_2021,young2021inertia}. 

\citeauthor*{kafiabad_savva_vanneste_2019} (\citeyear{kafiabad_savva_vanneste_2019}, hereafter \citetalias{kafiabad_savva_vanneste_2019}) used  multiscale asymptotics to show that the  wave-action  of linear IGWs propagating in a steady random geostrophic flow of much larger spatial scale evolves according to the diffusion equation
\begin{align}\label{diffusioneq}
    \partial_t a + \boldsymbol{c} \bcdot \bnabla_{\boldsymbol{x}} a
    = \bnabla_{\boldsymbol{k}} \bcdot \left( \mathsfbi{D} \bcdot \bnabla_{\boldsymbol{k}} a \right) + F.
\end{align}
Here $a(\boldsymbol{x}, \boldsymbol{k}, t)$ is the wave-action density in the $(\boldsymbol{x}, \boldsymbol{k})$ phase space, $\bk$ is the wavevector,  $\boldsymbol{c} = \bnabla_{\boldsymbol{k}} \omega$ is the intrinsic group velocity of IGWs, and $F(\boldsymbol{x}, \boldsymbol{k}, t)$ is a forcing term. The IGW intrinsic frequency
\begin{align}\label{frequency}
    \omega = (f^2 \cos^2 \theta + N^2 \sin^2 \theta)^{1/2},
\end{align}
with $f < N$ the Coriolis and buoyancy frequencies,
 depends on the angle $\theta$ between  $\boldsymbol{k}$ and the vertical. The $\boldsymbol{k}$-dependent diffusivity tensor  $\mathsfbi{D}$ is given in components by
\begin{align}\label{diffusiontensor}
    \mathsfi{D}_{ij} = k_m k_n \int_0^\infty \langle \partial_{x_i} U_n(\boldsymbol{x}) \partial_{x_j} U_m({\boldsymbol{x} - \boldsymbol{c}s})\rangle \, \text{d} s,
\end{align}
where $\langle \cdot \rangle$ denotes ensemble average and $\boldsymbol{U}$ is the flow velocity field, with prescribed homogeneous statistics.
A striking prediction of the diffusion equation \eqref{diffusioneq}
is that forced IGWs have a stationary spectrum scaling with wavenumber as $k^{-2}$, consistent with observed atmospheric mesoscale spectra \citep{gage-nastrom,lindborg99} and oceanic submesoscale spectra \citep{calli-ferr-jpo}. This provides support to the interpretation of the dynamics in these ranges as dominated by almost linear IGWs \citep{dewan1979,vanzandt,buhl-et-al14,call-et-al14,call-et-al16}. (The nature of the dynamics and level of nonlinearity  in the atmospheric mesoscales  is still a subject of debate; see \citet{li-linborg} and references therein for a contrasting view.)

\begin{figure}
    \centering
    \includegraphics[scale=1.]{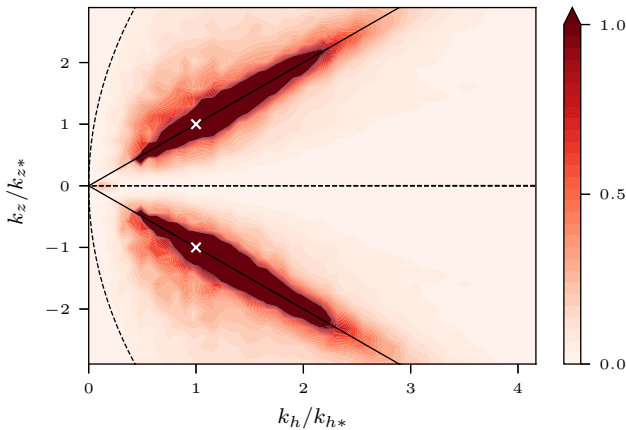}
    \caption{IGW energy spectrum $e$ as a function of the horizontal and vertical wavenumbers $(k_h,k_z)$ in the forced Boussinesq simulation described in \S\ref{comparison}. The wavenumbers are scaled by the forcing wavenumbers $(k_{h*},k_{z*})$ indicated by the white crosses. The cone corresponding to the forcing frequency is indicated by the solid lines. The energy $e$ is re-scaled by a characteristic value. The dashed lines indicate the boundary of the region of validity of the diffusion approximation (see appendix \ref{appA}).}
    \label{fig:contourdata}
\end{figure}

Crucially, the assumption of time-independent flow implies that the diffusivity tensor satisfies $\mathsfbi{D} \bcdot \boldsymbol{c} = 0$, as shown in \citetalias{kafiabad_savva_vanneste_2019}. Thus, noting $\mathsfbi{D}$ is symmetric, the diffusive flux $\mathsfbi{D} \bcdot \bnabla_{\boldsymbol{k}} a$ is perpendicular to $\boldsymbol{c} = \bnabla_{\boldsymbol{k}} \omega$ and hence the diffusion of wave action is restricted to a constant-frequency surface, namely a cone $\theta = \mathrm{const}$. This prediction is the direct consequence of the assumed linearity and  time independence. Simulations of the nonlinear Boussinesq equations reported by \citetalias{kafiabad_savva_vanneste_2019} nonetheless indicate that it applies to a good approximation to small-Rossby-number flows, because their time scale is asymptotically larger than the IGW propagation time scale. This is illustrated in figure \ref{fig:contourdata} which shows the result of a forced nonlinear Boussinesq simulation similar to  \citetalias{kafiabad_savva_vanneste_2019}'s  (see \S\ref{comparison} for details): the energy density in wavevector space is confined close to the constant-$\theta$ cone corresponding to the forcing frequency. 

However, \citet*{dong_buhler_smith_2020} suggest that the slow diffusion of wave action across constant-frequency surfaces that results from slow flow time dependence causes significant transfer of wave action from low to high frequency and demonstrate this for IGWs in rotating shallow water. The relevance of this result to three-dimensional IGWs is unclear. It is therefore an open question whether flow time dependence can radically alter the phenomenology of IGW diffusion by geostrophic turbulence, possibly on time scales much longer than the length of the simulations reported in \citetalias{kafiabad_savva_vanneste_2019} and in figure \ref{fig:contourdata}. 

We address this question in this paper by revisiting \citetalias{kafiabad_savva_vanneste_2019} to account for the slow time dependence of the geostrophic flow. Our starting point is the \citet{mccomas_bretherton_1977} diffusivity 
\begin{align}\label{diffusiontensortimedep}
    \mathsfi{D}_{ij} = k_m k_n \int_0^\infty \langle \partial_{x_i} U_n(\boldsymbol{x}, t) \partial_{x_j} U_m({\boldsymbol{x} - \boldsymbol{c}s}, t-s)\rangle \, \d s,
\end{align}
which applies to flows with arbitrary time dependence and was originally derived for wave--wave interactions in the induced diffusion regime. This diffusivity reduces to \eqref{diffusiontensor} in the time-independent case. (See \citet{dong_buhler_smith_2020} for a derivation using multiscale asymptotics.) Under the assumption of slow time dependence, 
encapsulated by a small parameter $\epsilon$ --  the ratio of the geostrophic flow velocity to the IGW group speed --
we approximate \eqref{diffusiontensortimedep} and solve the associated diffusion equation asymptotically to obtain  the equilibrium action distribution resulting from a steady single-frequency forcing. The results show that the action remains localised within an $O(\epsilon)$-thick boundary layer around the cone corresponding to the forcing frequency. This indicates that the diffusion of three-dimensional IGWs is largely unaffected by the slow time dependence of geostrophic turbulence.  In particular, the $k^{-2}$ equilibrium spectrum found by \citetalias{kafiabad_savva_vanneste_2019} can be recovered by integration of the solution across the boundary layer. We confirm the main theoretical predictions by comparison with a high-resolution simulation of the nonlinear Boussinesq equations as shown in figure \ref{fig:contourdata}.


 \section{Approximation of the diffusivity tensor}\label{sec:difftens}

In this section we approximate the diffusivity in \eqref{diffusiontensortimedep} taking advantage of the slow  time dependence of the geostrophic flow. 
Introducing the velocity correlation tensor $\mathsfi{\Pi}_{mn}(\boldsymbol{y},s) = \langle U_m(\boldsymbol{x} + \boldsymbol{y}, t + s)U_n(\boldsymbol{x}, t)\rangle$ we rewrite (\ref{diffusiontensortimedep}) as
\begin{align}\label{pi_diff}
    \mathsfi{D}_{ij} = -\frac{1}{2}k_m k_n \int_{-\infty}^\infty \frac{\partial^2 \mathsfi{\Pi}_{mn}}{\partial y_i \partial y_j} (\boldsymbol{c}s,s) \, \d s,
\end{align}
where we extend the integration range to $(-\infty, \infty)$ using that $k_m k_n \mathsfi{\Pi}_{mn}(-\boldsymbol{c}s,-s) = k_m k_n \mathsfi{\Pi}_{mn}(\boldsymbol{c}s,s)$. In terms of the wavevector--frequency  spectrum $\hat{\mathsfi{\Pi}}_{mn}$ defined via the Fourier transform
\begin{align}\label{fourierconvention}
    \mathsfi{\Pi}_{mn}(\boldsymbol{x}, t) = \int_{\mathbb{R}^4} \hat{\mathsfi{\Pi}}_{mn}(\boldsymbol{K}, \mathit{\Omega})\e^{\i(\boldsymbol{K}\bcdot \boldsymbol{x} -\mathit{\Omega}t)} \, \d \boldsymbol{K}\text{d} \mathit{\Omega}
\end{align}
this becomes
\begin{align}\label{precorrection}
    \mathsfi{D}_{ij} = \upi k_m k_n \int_{\mathbb{R}^4}K_i K_j \hat{\mathsfi{\Pi}}_{mn}(\boldsymbol{K}, \mathit{\Omega}) \delta(\boldsymbol{K} \bcdot \boldsymbol{c}- \mathit{\Omega}) \, \d \boldsymbol{K} \text{d} \mathit{\Omega}
\end{align}
on using $\int_\mathbb{R} \e^{\i(\boldsymbol{K}\bcdot \boldsymbol{c} -\mathit{\Omega})s} \, \d s = 2 \pi \delta( \boldsymbol{K} \bcdot \boldsymbol{c}- \mathit{\Omega})$.
Using the spherical polar coordinates $(k,\theta,\phi)$ for $\boldsymbol{k}$ and $(K,\mathit{\Theta},\mathit{\Phi})$ for $\boldsymbol{K}$ (lowercase symbols for IGW $\boldsymbol{k}$-space and uppercase symbols for  geostrophic flow $\boldsymbol{K}$-space), we compute
\begin{align}
    k_m k_n \hat{\mathsfi{\Pi}}_{mn}(\boldsymbol{K}, \mathit{\Omega}) = (k_1 K_2 - k_2 K_1)^2 E_\psi(\boldsymbol{K},\mathit{\Omega}) = 2 k^2 \sin^2 \theta \sin^2 \gamma E(\boldsymbol{K}, \mathit{\Omega}),
    \label{kmkn}
\end{align}
where $E_\psi$ is the spectrum of the streamfunction $\psi$ of the geostrophic flow (that is, the Fourier transform of $\langle \psi(\boldsymbol{x} + \boldsymbol{y}, t+s) \psi(\boldsymbol{x}, t) \rangle$),
\begin{equation}
\gamma = \mathit{\Phi} - \phi,
\label{gamma}
\end{equation}
and $E(\boldsymbol{K},\mathit{\Omega})$$= K^2 \sin^2 \mathit{\Theta} E_\psi/2$ is the geostrophic flow kinetic-energy spectrum.
Substituting \eqref{kmkn}--\eqref{gamma} into (\ref{precorrection}) yields
\begin{align}\label{precorrection2}
    \mathsfi{D}_{ij} = 2 \upi k^2 \sin^2 \theta \int_{\mathbb{R}^4}K_i K_j \sin^2 \gamma E(\boldsymbol{K}, \mathit{\Omega}) \delta(\boldsymbol{K} \bcdot \boldsymbol{c}- \mathit{\Omega}) \, \d \boldsymbol{K} \text{d} \mathit{\Omega}.
\end{align}

Following \citetalias{kafiabad_savva_vanneste_2019}, we assume that the flow is isotropic in the horizontal so that $E(\boldsymbol{K}, \mathit{\Omega})$ is independent of $\mathit{\Phi}$. In spherical polar coordinates, several components of $\mathsfbi{D}$ vanish. To see this, we replace $\mathit{\Phi}$ by $\gamma$ as an integration variable in \eqref{precorrection2} and express $\boldsymbol{K}$ in the local spherical basis $(\boldsymbol{e}_k, \boldsymbol{e}_\theta, \boldsymbol{e}_\phi)$ associated with $\boldsymbol{k}$. Thus we write
\begin{align}\label{K}
    \boldsymbol{K} = K \sin \mathit{\Theta}\left( (\sin \theta \cos \gamma + \cot \mathit{\Theta} \cos \theta) \boldsymbol{e}_k + (\cos \theta \cos \gamma - \cot \mathit{\Theta} \sin \theta)\boldsymbol{e}_\theta +\sin \gamma \boldsymbol{e}_\phi\right).
\end{align}
We can now use the parity of the integrand with respect to $\gamma$ in \eqref{precorrection2}, noting that $\delta(\boldsymbol{K} \bcdot \boldsymbol{c}- \mathit{\Omega})$ is even since  $\boldsymbol{c} = \bnabla_{\boldsymbol{k}} \omega(\theta)$ implies that $\boldsymbol{c} \parallel \boldsymbol{e}_\theta$ hence
$\boldsymbol{K} \bcdot \boldsymbol{c} = \boldsymbol{e}_\theta \bcdot \boldsymbol{K} \,  c$. The parity of the integrands giving the components $\mathsfi{D}_{kk} = \boldsymbol{e}_k \bcdot \mathsfbi{D} \bcdot  \boldsymbol{e}_k$, etc.\ of $\mathsfbi{D}$ is then 
determined by the parity of pairwise products of  $\boldsymbol{e}_k \bcdot \boldsymbol{K}$, $\boldsymbol{e}_\theta \bcdot \boldsymbol{K}$ and $\boldsymbol{e}_\phi \bcdot \boldsymbol{K}$.  We conclude from this that the only non-zero components of $\mathsfbi{D}$ are $\mathsfi{D}_{kk}, \mathsfi{D}_{k \theta} = \mathsfi{D}_{\theta k}, \mathsfi{D}_{\theta \theta},$ and $\mathsfi{D}_{\phi \phi}$. Thus, diffusion in the azimuthal direction depends only on azimuthal gradients of action and is decoupled from the $k$ and $\theta$ directions.

We now restrict our attention to flows that are slowly time dependent in the sense that their typical frequencies $\mathit{\Omega}$ and wavevectors $\boldsymbol{K}$ satisfy $\mathit{\Omega} \ll \boldsymbol{K} \bcdot \boldsymbol{c}$. 
For realistic turbulent flows, $\mathit{\Omega} \sim UK$, hence this condition is equivalent to the condition $U \ll c$ that underpins the diffusion approximation (\ref{diffusioneq}), the limitation of which is discussed in appendix \ref{appA}.
 To make the smallness of $\mathit{\Omega}$ relative to $\boldsymbol{K} \bcdot \boldsymbol{c}$ explicit, we introduce a  bookkeeping parameter $\epsilon \ll 1$ to mark out asymptotically small terms. The delta function $\delta(\boldsymbol{K} \bcdot \boldsymbol{c} - \mathit{\Omega})$ in (\ref{precorrection2}) becomes  $\delta(\boldsymbol{K} \bcdot \boldsymbol{c} - \epsilon \mathit{\Omega})$ and can be expanded as
 \begin{align}\label{expansion}
     \delta(\boldsymbol{K} \bcdot \boldsymbol{c} - \epsilon \mathit{\Omega}) &= \delta(\boldsymbol{K} \bcdot \boldsymbol{c}) - \epsilon \delta'(\boldsymbol{K} \bcdot \boldsymbol{c}) \mathit{\Omega} + \epsilon^2 \delta''(\boldsymbol{K} \bcdot \boldsymbol{c}) \mathit{\Omega}^2/2 + O(\epsilon^3).
 \end{align}
 Using this alongside the evenness of $E(\boldsymbol{K}, \mathit{\Omega})$ in $\mathit{\Omega}$ leads to the approximation
 \begin{align}
      \mathsfbi{D} &= \mathsfbi{D}^{(0)} + \epsilon^2 \mathsfbi{D}^{(1)} +O(\epsilon^4).
\end{align}
 Here 
 \begin{align}
     \mathsfi{D}^{(0)}_{ij} &= 2 \upi k^2 \sin^2 \theta \int_{\mathbb{R}^3}K_i K_j \sin^2 \gamma E(\boldsymbol{K}) \delta(\boldsymbol{K} \bcdot \boldsymbol{c}) \, \d \boldsymbol{K},
     \label{D0}
\end{align}
where $E(\boldsymbol{K})$ is the geostrophic flow kinetic energy spectrum marginalised over frequencies, 
 recovers the diffusivity of time-independent flows obtained by \citetalias{kafiabad_savva_vanneste_2019} (up to a factor  $(2\upi)^3$ corresponding to a different Fourier transform convention, see \eqref{fourierconvention}). 
 For a horizontally isotropic geostrophic flow,  $\mathsfi{D}^{(0)}_{ij}$ has two non-zero components in spherical polar coordinates, namely
\begin{subequations} \label{KSVdiffusivity}
\begin{align}
    \mathsfi{D}^{(0)}_{kk} &= \frac{4 \upi k^3 \omega \sin^2\theta}{(N^2 - f^2) |\cos^5 \theta|}\int_{-\infty}^{\infty} \int_{\theta}^{\upi - \theta} K^3 \cos^2\mathit{\Theta} (\cot^2\theta - \cot^2 \mathit{\Theta})^{1/2}E(K,\mathit{\Theta}) \, \d K \text{d} \mathit{\Theta}, \label{diffusiontensork}\\
    \mathsfi{D}^{(0)}_{\phi\phi} &= \frac{4 \upi k^3 \omega \sin^4\theta}{(N^2 - f^2) |\cos^5 \theta|}\int_{-\infty}^{\infty} \int_{\theta}^{\upi - \theta} K^3 \sin^2\mathit{\Theta} (\cot^2\theta - \cot^2 \mathit{\Theta})^{3/2}E(K,\mathit{\Theta}) \, \d K \text{d} \mathit{\Theta}. \label{diffusiontensorphi}
\end{align}
\end{subequations}
(These equations are  (A 13) in \citetalias{kafiabad_savva_vanneste_2019}, up to the $(2\upi)^3$ factor and a typographical correction in the lower limit of $\theta$.)

The leading-order correction to \eqref{D0} induced by the slow flow time dependence is
\begin{align}\label{correctionintegral}
  \mathsfi{D}^{(1)}_{ij} = \upi k^2 \sin^2 \theta \int_{\mathbb{R}^3}K_i K_j \sin^2 \gamma A(\boldsymbol{K}) \delta''(\boldsymbol{K} \bcdot \boldsymbol{c}) \, \d \boldsymbol{K},
\end{align}
and depends on the geostrophic-flow acceleration spectrum
\begin{align}
    A(\boldsymbol{K}) = \int_{\mathbb{R}} E(\boldsymbol{K}, \mathit{\Omega})\mathit{\Omega}^2 \, \d \mathit{\Omega},
\end{align}
a natural measure of the flow's unsteadiness.

It turns out that the only dynamically significant component of $\mathsfbi{D}^{(1)}$ is $\mathsfi{D}^{(1)}_{\theta \theta}$, corresponding to across-cone diffusion, on which we now concentrate. Contracting \eqref{correctionintegral} twice with $\boldsymbol{e}_\theta = \boldsymbol{c}/c$, we obtain 
\begin{align}\label{dthetatheta}
    \mathsfi{D}^{(1)}_{\theta \theta}= \frac{\boldsymbol{c} \bcdot \mathsfbi{D}^{(1)} \bcdot \boldsymbol{c}}{c^2} = \frac{\upi k^2 \sin^2 \theta}{c^2} \int_{\mathbb{R}^3} (\boldsymbol{K}\bcdot \boldsymbol{c})^2 \sin^2 \gamma A(\boldsymbol{K}) \delta''(\boldsymbol{K} \bcdot \boldsymbol{c})\, \d \boldsymbol{K}.
\end{align}
Noting that
\begin{align}
    \int_\mathbb{R} x^2f(x)\delta''(x)\, \d x = 2 \int_\mathbb{R} f(x)\delta(x)\, \d x
\end{align}
for any smooth $f(x)$ reduces (\ref{dthetatheta}) to
\begin{align}\label{dthetatheta2}
    \mathsfi{D}^{(1)}_{\theta \theta} = \frac{2 \upi k^2 \sin^2 \theta}{c^2} \int_{\mathbb{R}^3} \sin^2 \gamma A(\boldsymbol{K}) \delta(\boldsymbol{K} \bcdot \boldsymbol{c})\, \d \boldsymbol{K}.
\end{align}
Representing $\boldsymbol{K}$ in the polar spherical coordinates $(K,\mathit{\Theta},{\gamma})$ and expanding $\boldsymbol{K} \bcdot \boldsymbol{c}$ using (\ref{K}) gives
\begin{align}
    \mathsfi{D}^{(1)}_{\theta \theta} = \frac{2 \upi k^2 \sin^2 \theta}{c^2} \int_{0}^{\infty} \d K\int_{0}^{\pi} \text{d} \mathit{\Theta} \int_{-\pi}^{\pi} \text{d} \gamma \, &K^2 \sin \mathit{\Theta} \sin^2 \gamma A(K,\mathit{\Theta}) \label{dthetatheta3} \\
    &\times \delta\left(K c \sin \mathit{\Theta} \cos\theta \left(\cos \gamma - \frac{\cot \mathit{\Theta}}{\cot \theta}\right)\right), \nonumber 
\end{align}
where we use horizontal isotropy to write $A(\boldsymbol{K})= A(K,\mathit{\Theta}) $. 
Under the change of variable $\zeta = \cos \gamma$ this simplifies into 
\begin{align}
    \mathsfi{D}^{(1)}_{\theta \theta} = \frac{4 \upi k^2 \sin^2 \theta}{c^3 |\cos \theta|} \int_{0}^{\infty}\text{d} K\int_{0}^{\pi} \text{d} \mathit{\Theta} \int_{-
    1}^{1} \text{d} \zeta \, K (1-\zeta^2)^{1/2} A(K,\mathit{\Theta})
    \delta\left(\zeta - \frac{\cot \mathit{\Theta}}{\cot \theta}\right), \label{dthetatheta5}
\end{align}
where the factor of 2 arises from the evenness of $\cos \gamma$. Only values of $\mathit{\Theta}$ for which $|\cot \mathit{\Theta}/\cot \theta| < 1$ contribute to the integral, which reduces the integration range  to $(\theta, \upi - \theta)$. Integrating over $\zeta$ then yields
\begin{align}
    \mathsfi{D}^{(1)}_{\theta \theta} = \frac{4 \upi k^2 \sin^2 \theta}{c^3 |\cos \theta|} \int_{0}^{\infty}\text{d} K\int_{\theta}^{\pi - \theta} \text{d} \mathit{\Theta} \, K \left(1-\left(\frac{\cot \mathit{\Theta}}{\cot \theta}\right)^2\right)^{1/2} A(K, \mathit{\Theta}).\label{dthetatheta6}
\end{align}
Substituting in 
\begin{align}\label{groupvel}
c = |\bnabla_{\boldsymbol{k}} \omega| = \frac{1}{k}\partial_\theta (f^2 \cos^2 \theta + N^2 \sin^2 \theta)^{1/2} = \frac{(N^2 - f^2)|\sin\theta \cos\theta|}{\omega k},
\end{align}
and rearranging gives the final form 
\begin{align}
    \mathsfi{D}^{(1)}_{\theta \theta} = \frac{4 \upi \omega^3 k^5}{(N^2 - f^2)^3 |\cos^5 \theta| } \int_{0}^{\infty}\int_{\theta}^{\pi - \theta}  K (\cot^2\theta- \cot^2 \mathit{\Theta})^{1/2} A(K,\mathit{\Theta})\, \text{d} K\text{d} \mathit{\Theta}.\label{dthetathetafinal}
\end{align}

In summary, the diffusivity with time-dependent geostrophic flow has 3 significant components: $\mathsfi{D}_{kk} = \mathsfi{D}^{(0)}_{kk}$ and $\mathsfi{D}_{\phi \phi} = \mathsfi{D}^{(0)}_{\phi \phi}$ given by \eqref{KSVdiffusivity} and dependent on the energy spectrum of the geostrophic flow, and $\mathsfi{D}_{\theta \theta} = \epsilon^2 \mathsfi{D}^{(1)}_{\theta \theta}$ given by (\ref{dthetathetafinal}) and dependent on the flow acceleration spectrum. The small, non-zero $\mathsfi{D}_{\theta \theta}$  for non-vanishing flow acceleration captures the weak cross-cone diffusion pointed out by \citet{dong_buhler_smith_2020}.

\section{Equilibrium spectrum} \label{sec:eq}

\subsection{Solution of the steady diffusion equation}


We now focus on the response to the spatially homogeneous, azimuthally isotropic steady forcing 
\begin{align}\label{Forcingdef}
F(\boldsymbol{k}) = \delta(k-k_*) \delta(\theta - \theta_*)
\end{align}
corresponding to a single IGW frequency. (The response to a forcing with arbitrary dependence on $k$ and $\theta$ can be obtained by integration.)
We aim to show that the action density reaches an equilibrium $a(k,\theta)$ that is localised near $\theta = \theta_*$ -- in other words, that the frequencies remain close to the forcing frequency for all time. 
This is in contrast with the two-dimensional case of \citet{dong_buhler_smith_2020} for which no such localised equilibrium exists. 

For ease of interpretation, we replace the action density by the energy density
$e(k,\theta) = 2\upi k^2 \sin \theta \omega a(k, \theta)$, such that $e \, \text{d}k \text{d}\theta$ is the energy contained in the box $[k, k + \text{d}k]$ and $[\theta, \theta + \text{d} \theta]$. Eq.\ \eqref{diffusioneq} then reduces to
\begin{align}\nonumber
    \partial_k\left( k^2 \left(\mathsfi{D}^{(0)}_{kk} + \epsilon^2 \mathsfi{D}^{(1)}_{kk} \right) \partial_k \frac{e}{k^2} + \epsilon^2 \frac{\sin\theta \omega}{k} \mathsfi{D}^{(1)}_{k \theta} \partial_\theta \frac{e}{\sin\theta \omega} \right)&\\
    + \epsilon^2 \omega k \partial_\theta \left(\frac{1}{\omega} \mathsfi{D}^{(1)}_{k \theta}\partial_k \frac{e}{k^2} + \frac{\sin \theta}{k^3} \mathsfi{D}^{(1)}_{\theta \theta} \partial_\theta \frac{e}{\sin \theta \omega} \right)&
    = -\delta(k-k_*) \delta(\theta - \theta_*), \label{expansion2}
\end{align}
where we ignore unimportant prefactors on the right-hand side. 
We seek solutions localised in $\theta$ in a boundary layer of thickness $\epsilon$ around $\theta_*$, assuming
\begin{equation} \label{sigma}
\sigma = (\theta-\theta_*)/\epsilon = O(1).
\end{equation}
To leading-order in $\epsilon$, \eqref{expansion2} reduces to
\begin{align}
    \partial_k\left( k^2 \mathsfi{D}^{(0)}_{kk}(\theta_*) \partial_k \frac{e}{k^2}\right) + \frac{1}{k^2} \mathsfi{D}^{(1)}_{\theta \theta}(\theta_*) \partial_{\sigma \sigma} e = -\delta(k - k_*)\delta(\sigma),\label{expansion3}
\end{align}
ignoring again a prefactor on the right-hand side (in this case $1/\epsilon$).
Note that $\mathsfi{D}^{(1)}_{\theta \theta}$ is the only  correction to the diffusivity tensor induced by flow time dependence that appears in \eqref{expansion3}. (This also applies to anisotropic IGWs in the sense that $\partial_\phi a \not= 0$.) This correction appears at leading order, even though the corresponding diffusivity $\epsilon^2 \mathsfi{D}^{(1)}_{\theta \theta}$ is small, because of the large gradients in $\theta$ of the solution.



We make the dependence on $k$ of the diffusivity components $\mathsfi{D}^{(0)}_{kk}$ and 
$\mathsfi{D}^{(1)}_{\theta \theta}$  in (\ref{diffusiontensork}) and (\ref{dthetathetafinal}) explicit by writing
\begin{align}
\mathsfi{D}^{(0)}_{kk} = Q(\theta) k^3 \quad \textrm{and} \quad \mathsfi{D}^{(1)}_{\theta \theta} = R(\theta) k^5.
\label{RQ}
\end{align} 
Under the change of variables
\begin{align}\label{variablechange}
    e = \bar{e}/(Q_*R_*)^{1/2} \quad \text{and} \quad \sigma = \bar{\sigma} (R_{*}/Q_*)^{1/2},
\end{align}
where $Q_* = Q(\theta_*)$ and $R_* = R(\theta_*)$, \eqref{expansion3}  becomes
\begin{align}\label{finaldiffusioneq}
    k^3 \partial_{kk}\bar{e} + k^2 \partial_k \bar{e} - 4 k \bar{e} + k^3 \partial_{\bar{\sigma}\bar{\sigma}} \bar{e} = - \delta(k - k_*)\delta(\bar{\sigma}).
\end{align}
In the following, we drop the overbars for simplicity. 

We now solve the re-scaled  problem \eqref{finaldiffusioneq}. Taking  a Fourier transform in $\sigma$, with $l$ the corresponding Fourier variable, we find
\begin{align}\label{fouriereq}
    k^3 \partial_{kk}\hat{e} + k^2\partial_{k}\hat{e} -4k\hat{e} - k^3 l^2 \hat{e} = -\frac{\delta(k-k_*)}{2\upi},
\end{align}
where the hat denotes the Fourier transform. The solution to the homogeneous problem can be written in terms of modified Bessel functions \citep[][Ch.\ 10]{NIST:DLMF}, leading to the piecewise expression
\begin{align}\label{generalpiecewise}
    \hat{e}(k,l) =\left\{
    \begin{array}{ll}
      A(l)\mathrm{I}_2(|l|k) + B(l)\mathrm{K}_2(|l|k) & \textrm{for} \ 0 < k < k_* \\
      C(l)\mathrm{I}_2(|l|k) + D(l)\mathrm{K}_2(|l|k)   & \textrm{for} \  k > k_*
    \end{array} \right.,
\end{align}
where $\mathrm{I}$ and $\mathrm{K}$ are modified Bessel functions of the first and second kind, and $A, B, C$ and $D$ are so far arbitrary functions of $l$. These  functions are determined by the boundary and jump conditions. Finiteness as $k \to 0$ and $k \to \infty$ requires that $B(l)=C(l)=0$. Imposing continuity at $k_*$ and the jump $[\partial_k \hat e]_{k_*^-}^{k_*+} = -1/(2 \pi k_*^3)$ then gives 
\begin{align}
    \begin{pmatrix}
    A\\
    D
    \end{pmatrix}
    =
    \frac{1}{2 \upi \mathcal{W}\{\mathrm{K}_2(|l|k_*), \mathrm{I}_2(|l|k_*)\}|l|k_*^3}
    \begin{pmatrix}
    \mathrm{K}_2(|l|k_*)\\
    \mathrm{I}_2(|l|k_*)
    \end{pmatrix} =
    \frac{1}{2\upi k_*^2}
    \begin{pmatrix}
    \mathrm{K}_2(|l|k_*)\\
    \mathrm{I}_2(|l|k_*)
    \end{pmatrix},
\end{align}
where $\mathcal{W}$ is the Wronskian and we use that
$\mathcal{W}\{\mathrm{K}_2(z), \mathrm{I}_2(z)\} = {1}/{z}$
\citep[][Eq.\ (10.28.2)]{NIST:DLMF}.
Hence the solution in Fourier space is
\begin{align}\label{afterBCs}
    \hat{e}(k,l) =\frac{1}{2 \upi k_*^2}  \times \left\{
    \begin{array}{ll}
      \mathrm{K}_2(|l|k_*)\mathrm{I}_2(|l|k) & \textrm{for} \ 0 < k < k_* \\
     \mathrm{I}_2(|l|k_*)\mathrm{K}_2(|l|k)         &  \textrm{for} \ k > k_*
    \end{array} \right..
\end{align}

We invert the Fourier transform. As $\hat{e}$ is symmetric in $l$, the inverse of \eqref{afterBCs} is
\begin{align}\label{FTcases}
    e(k, \sigma)  = \frac{1}{\upi k_*^2} \times \left\{
    \begin{array}{ll}
      \int_{0}^{\infty}\mathrm{K}_2(lk_*)\mathrm{I}_2(lk)\cos(\sigma l) \, \d l & \textrm{for} \ 0 < k < k_* \\
     \int_{0}^{\infty}\mathrm{I}_2(lk_*)\mathrm{K}_2(lk)\cos(\sigma l) \, \d l         & \textrm{for} \  k > k_*
    \end{array} \right. .
\end{align}
This can be evaluated exactly using Eq.\ (4), \S 6.672 of \citet[]{2014637},
\begin{align}
    \int_0^\infty \mathrm{K}_{v}(ax) \mathrm{I}_{v}(bx) \cos(cx) \, \d x = \frac{1}{2(ab)^{1/2}}\mathrm{Q}_{v - 1/2} \left(\frac{a^2 + b^2 + c^2}{2ab} \right),
\end{align}
which holds providing that $\Real(a) > |\Real(b)|$ and $\Real(v)> -1/2 $. Here, $\mathrm{Q}_{v - 1/2}$ is the  Legendre function of the second kind \citep[][Ch.\ 14]{NIST:DLMF}. Clearly, the condition on $v$ is satisfied for (\ref{FTcases}).  For $0<k<k_*$, $a = k_* > k = b$; for $k > k_*$, $a = k > k_* = b$. Therefore, the condition on $a$ and $b$ also holds. Due to the symmetry of the solution under exchanges of $a$ and $b$, both integrals in \eqref{FTcases} are equivalent, leading to
\begin{align}\label{e_prescaling}
    e(k, \sigma) = \frac{1}{2\upi k_{*}^{5/2}k^{1/2}}\mathrm{Q}_{3/2} \left(\frac{k_*^{2} + k^{2} + \sigma^2}{2k_* k} \right). 
\end{align}

Eq.\ \eqref{e_prescaling} is the main result of the paper. It gives the form for the equilibrium distribution of IGW energy forced at a single wavenumber and frequency, accounting for the time dependence of the turbulence. Since $\mathrm{Q}_{3/2}$ decays rapidly as its argument increases, \eqref{e_prescaling} shows that the IGW energy is localised within an $O(\epsilon)$ layer around the constant frequency cone $\theta = \theta_*$ (recall \eqref{sigma}). Note that $e(k,\sigma)$ has a mild, logarithmic singularity as $\sigma \to 0$ for $k=k_*$.

\begin{figure}
    \centering
    \includegraphics[trim = {.7cm 0cm -.7cm 0cm}]{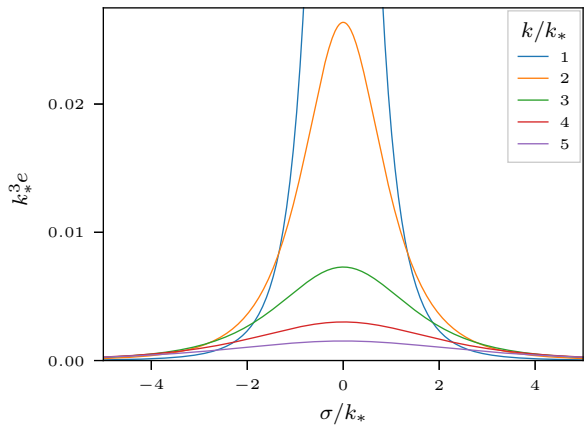}
    \caption{IGW energy spectrum $e$ in \eqref{e_prescaling} scaled by $k_*^3$ as a function of the scaled angle $\sigma/k_*$ for a few values of non-dimensionalised total wavenumber $k/k_*$.}
    \label{fig:exact}
\end{figure}

\begin{figure}
    \centering
    \includegraphics{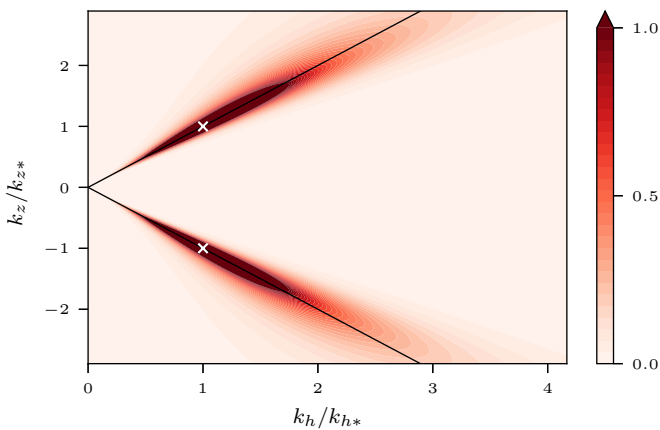}
    \caption{IGW energy spectrum $e$ in \eqref{e_prescaling} as a function of horizontal and vertical wavenumbers $(k_h, k_z)$. The wavenumbers are scaled by the forcing wavenumbers $(k_{h*}, k_{z*})$ indicated by the white crosses. The cone corresponding to the forcing frequency is indicated by the solid lines. The parameters $Q_*$ and $R_*$ are chosen to match the simulation results in \S\ref{comparison} (cf.\ figure \ref{fig:contourdata}).}
    \label{fig:contourexact}
\end{figure}

We illustrate the form of the energy spectrum predicted by \eqref{e_prescaling} in figure \ref{fig:exact}. Here, $e$ scaled by $k_*^3$ is plotted against the scaled angle $\sigma/k_*$ for a few values of non-dimensionalised total wavenumber $k/k_*$. In figure \ref{fig:contourexact}, $e$ is shown as a function of horizontal and vertical wavenumber and is scaled to approximately match the energy level of the simulation in \S\ref{comparison}. The value of $R_*/Q_*$ is also required for figure \ref{fig:contourexact} and is chosen to match simulation results.

A useful approximation to \eqref{e_prescaling} is obtained from the asymptotics of the Legendre function for large argument:
\begin{align}\label{generallimit}
    e(k, \sigma) \sim 
    \frac{3}{16} k^{-3}\left(1 + \frac{k_{*}^{2} + \sigma^2}{k^2}\right)^{-5/2},
\end{align}
which applies for $k \to 0$, $k \to \infty$ or $\sigma \to \infty$. In particular, it makes it possible to characterise the angular localisation of the energy by the power law
\begin{align}\label{sigma_5}
    e(k, \sigma) \sim \frac{3}{16} \frac{k^2}{\sigma^5} \quad \textrm{as} \quad \sigma \rightarrow \infty.
\end{align}
Eq.\ \eqref{generallimit} further shows that at fixed $\sigma$, that is,  at fixed angle $\theta$ or frequency, $e(k,\sigma) \propto k^{-3}$ as $k \to \infty$, and  $e(k,\sigma) \propto k^{2}$ for $k \ll k_*$.

Another limit of interest deduced from \eqref{generallimit} is  
\begin{align} \label{easym}
    e(k, \sigma) \sim \frac{3}{16} k^{-3}\left(1 + \left(\frac{\sigma}{k}\right)^2\right)^{-5/2} \quad  \textrm{as} \quad k \to \infty, \ \sigma/k=O(1),
\end{align}
which shows that the spectrum broadens in $\sigma$ like $k$. Consequently, integration of \eqref{generallimit} across angles results in a spectrum decaying like $k^{-2}$. In fact, the integrated spectrum is exactly proportional to $k^{-2}$ for $k > k_*$: indeed, integration of \eqref{finaldiffusioneq} with respect to $\bar{\sigma}$ recovers the equation found by \citetalias{kafiabad_savva_vanneste_2019} for time-independent flows, with solution proportional to $k^{-2}$ for $k > k_*$ and $k^2$ for $k< k_*$. 

In dimensional terms, the thickness of the boundary layer around the cone is proportional to the square root of the ratio $R_*/Q_*$ (see \eqref{variablechange}), which roughly amounts to the ratio of the flow acceleration to its energy, and can be interpreted as the relevant flow frequency. This increases when the flow becomes more transient resulting in a thicker boundary layer.


\subsection{Comparison with Boussinesq simulations}\label{comparison}

We compare the analytical prediction \eqref{e_prescaling}  with the results of a high-resolution three-dimensional Boussinesq simulation. We solve the non-hydrostatic Boussinesq equations using a de-aliased pseudospectral code adopted from that in \citet{waite2006transition}. A third-order Adams--Bashforth scheme with timestep $0.0044/f$, is employed for time integration. The triply-periodic domain $\left[0, 2 \pi\right]^2 \times \left[0, 2 \pi f/N\right]$ is discretised with $2304^3$ grid points. A hyperdissipation of the form $\nu_h (\partial_x^2+\partial_y^2)^4 + \nu_z \partial_{z}^8$, with $\nu_h = 7.8 \times 10^{-23}$ and $\nu_z = 7.1 \times 10^{-35}$ (in dimensionless units, with the domain size as reference length and $f^{-1}$ as reference time) is implemented in the momentum and buoyancy equations. We take $N/f=32$, a representative value of mid-depth ocean stratification. We initialise the simulation with a fully-developed geostrophic turbulent flow, which is the output of a decaying quasigeostrophic model with the initial energy spectrum proportional to $\exp{(\textstyle -(((K_h^2 + f^2 K_z^2/N^2)^{1/2}-24)/10)^2)}$. This model is run until the energy spectrum fills the spectral space, peaking at $K_h = 4$ and scaling approximately as $K_h^{-3}$ and $K_z^{-3}$.  The flow parameters are selected such that the Rossby number based on the vertical vorticity $\zeta$ is ${\rm{Ro}} = \langle  \zeta^2 \rangle^{1/2}/f = 0.11$.   Throughout the simulation, an Ornstein–Uhlenbeck forcing with short correlation time (3 timesteps) is applied
to the linear wave modes with $(k_{h*},k_{z*})=(12,221)$  corresponding to the fixed IGW frequency of $2f$. This relatively low frequency is chosen so that the aspect ratio of the IGWs  is similar to the aspect ratio $N/f$ of the geostrophic flow and thus the IGWs are well resolved with the anisotropic grid we use. 
The simulation is performed until $t = 160/f$ by which time the statistics are approximately stationary. We separate IGWs from the mean flow (both for forcing and extracting energy spectra) using the normal-mode decomposition of \cite{bartello95}.

\begin{figure}
    \centering
    \includegraphics[scale=1.]{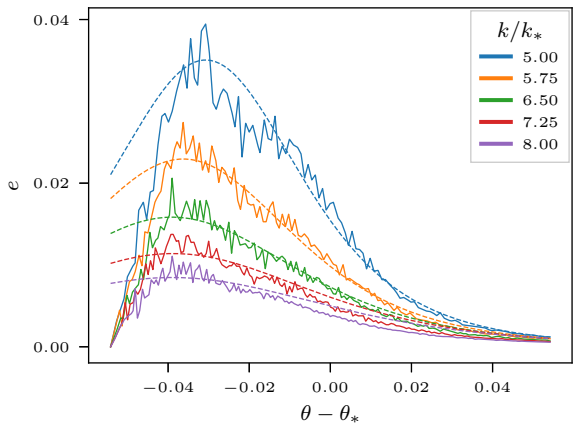}
    \caption{IGW energy spectrum $e$ vs $\theta-\theta_*$ for several values of $k/k_*$: comparison between simulation results (solid lines) and analytical prediction \eqref{e_prescaling} (dashed lines). The scalings of $e$ and $\sigma \propto \theta - \theta_*$ are chosen for the analytical prediction to best match the simulation data (scaling $\sigma$ corresponds to estimating $(R_*/Q_*)^{1/2}$, see  \eqref{sigma} and \eqref{variablechange}).}
    \label{fig:fit}
\end{figure}

 We compare the functional form implied by \eqref{e_prescaling} to the spectrum $e(k_h,k_z)$ obtained in the simulation. This involves fitting two parameters, one that fixes the scale of $e$ and corresponds to strength of the forcing, and the other that fixes the scale of $\sigma$ and corresponds to $(R_*/Q_*)^{1/2}$ (see \eqref{variablechange}).  We estimate these two parameters by matching the simulation spectrum as a function of $\theta - \theta_*$ for  $k \gtrsim 5 k_*$ as shown in figure \ref{fig:fit}. These values of $k$ are large enough for the perturbation induced by the non-ideal nature of the forcing in the simulation to be negligible, and for discretisation effects to play only a minor role.
  A difficulty, evident in figure \ref{fig:fit}, is that the simulation spectrum is not symmetric. 
  We attribute this to an edge effect caused by the proximity of the IGW frequency $\omega = 2 f$ to the minimum allowable frequency $\omega = f$,  and to the breakdown of the diffusion approximation when $\omega$ is close to $f$ (see appendix \ref{appA} for details). (The forcing frequency cone has a small opening angle, $\theta_* = \tan^{-1} (k_{h*}/k_{z*}) \approx 3^\circ$, a feature obscured by the anisotropic scaling of the axes in figures 
\ref{fig:contourdata} and \ref{fig:contourexact}.) We therefore carry out the parameter fitting based on the parts of the curves in figure \ref{fig:fit} right of their maxima. We further allow for an offset of $\theta - \theta_*$, likely the result of the coarse discretisation of the wavevector in the forcing region. 

\begin{figure}
\begin{center}
\begin{tabular}{cc}
(a) & (b) \\
  \includegraphics[trim={.2cm 0cm 0cm 0cm},scale=1.]{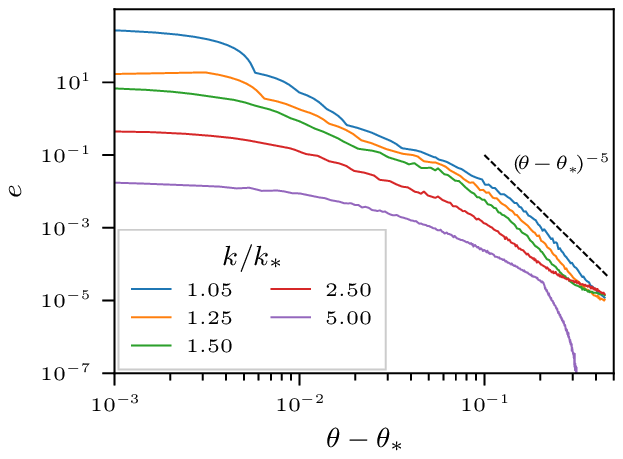} &
  \includegraphics[trim={0cm 0cm .2cm 0cm},scale=1.]{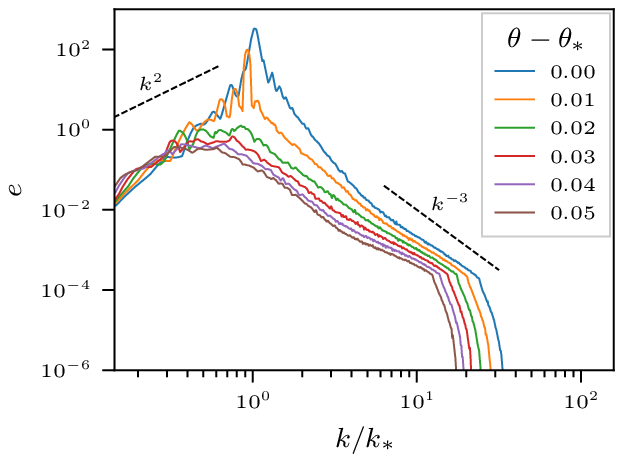}
\end{tabular}
  \caption{IGW energy spectrum $e$ from simulation data in log--log coordinates: (a) as a function of $\theta-\theta_*$ for several values of $k/k_*$, and (b) as a function of $k/k_*$ for several values of $\theta-\theta_*$. Predicted power laws are indicated by dashed lines.} 
\label{fig:loglog}

\end{center}
\end{figure}

The prediction of \eqref{e_prescaling} with the two fitted parameters is shown by the dashed curves in figure \ref{fig:fit}. 
The agreement with the numerical results is good: \eqref{e_prescaling} captures the localisation of $e$ and the  general form of its decrease with $\theta - \theta_*$ at different values of $k$. (We emphasise that the same two parameters are used for all the curves.) A complementary view is provided by figure \ref{fig:loglog} which shows $e$ obtained in the simulation as a function of $\theta - \theta_*$ (panel (a)) and of $k/k_*$ (panel (b)) in log--log coordinates. The power laws $\sigma^{-5}$ (equivalent to $(\theta - \theta_*)^{-5}$), $k^{-3}$ and $k^2$ derived in \eqref{generallimit}--\eqref{sigma_5} from \eqref{e_prescaling} are shown in their range of expected validity. The $\sigma^{-5}$ and $k^{-3}$ power laws are consistent with the data albeit over a limited wavenumber range. We regard this as a reasonable match given the difficulties in capturing such rapid decay in a numerical model, and the pollution by the forcing. The $k^{2}$ power law is a poorer match. 
This is be expected since the spatial scale-separation assumption between IGWs and geostrophic flow that underpins the diffusion equation \eqref{diffusioneq} is not satisfied for  wavenumbers smaller than the forcing  wavenumber. The numerical spectrum for small $k$ is also strongly affected by discretisation effects.
Note that the abrupt drop in the tail of spectra in figure \ref{fig:loglog}(b) comes from the truncation of data due to storage limitation; the total energy spectrum shows a smooth transition to dissipation range (not shown).

Overall, the simulation results compare  as well with \eqref{e_prescaling} as can be expected given the numerical challenges posed by the finite resolution, non-ideal forcing and an IGW signal that has both low amplitude and decreases rapidly with $k$ and $\theta - \theta_*$. We  note that it is in principle possible to compute the scaling parameter $(R_*/Q_*)^{1/2}$ from simulation data using the explicit expressions for $R_*$ and $Q_*$  deduced from (\ref{diffusiontensork}),  (\ref{dthetathetafinal}) and (\ref{RQ}). \citetalias{kafiabad_savva_vanneste_2019} evaluate $Q_*$ based on the energy spectrum of the geostrophic flow they estimate from simulation data. An analogous evaluation of $R_*$ requires the acceleration spectrum of the geostrophic flow. We leave this computation for future work.

\section{Discussion}

This paper is part of a sequence of works that apply techniques of waves in random media to address the role of the geostrophic flow in shaping the energy distribution of atmospheric and oceanic inertia-gravity waves \citep{dani-v16,savv-vann18,kafiabad_savva_vanneste_2019,savva_kafiabad_vanneste_2021}. Their main assumption is that the flow is weak enough to be regarded as a small perturbation to what would otherwise be IGWs propagating in a medium at rest. The perturbation,  physically transport and refraction, can be interpreted as arising from resonant triadic interactions involving two IGW modes and a geostrophic (or vortical) mode -- these are known as `catalytic interactions' in recognition of the fact that the geostrophic mode is left unaffected \citep{lelo-rile,bartello95}. 
The present paper further assumes that the IGWs have spatial scales much smaller than the flow scales. In this case, the impact of the flow, modelled as a random field, on the IGWs is a diffusion of wave action in wavevector space. (This is the induced diffusion regime considered by \citet{mccomas_bretherton_1977} in the context of wave--wave interactions.) \citetalias{kafiabad_savva_vanneste_2019} examined this process in some detail and showed, in particular, that it leads to IGW characteristics such as a $k^{-2}$ stationary spectrum that are consistent with atmospheric and oceanic observations. 

To obtain these results, \citetalias{kafiabad_savva_vanneste_2019} treated the geostrophic flow as time independent, on the grounds that it evolves on a time scale much longer than the IGW periods. With this assumption, the geostrophic mode has a zero frequency. The (resonant) catalytic interactions therefore involve two IGW modes with exactly the same frequency, and wave action exchanges are restricted to a constant-frequency surface in wavevector space. Here, we revisit this assumption by taking the geostrophic flow to be slowly evolving. 
In this case, the catalytic interactions are between a low frequency geostrophic mode and two IGWs with slightly different frequencies, and action  diffuses across the constant-frequency surface. The question is therefore whether this leads to qualitative changes in the statistics of IGWs, for instance by enabling IGW frequencies to diffuse freely and wave action to spread unimpeded across wavevector space (as was recently shown to be the case for two-dimensional waves by \citet{dong_buhler_smith_2020}). The answer is no: we show that the stationary spectrum established by forcing single-frequency IGWs  is localised within a boundary layer close to the cone of constant frequency associated with the forcing. Thus, even in the infinite-time limit corresponding to this stationary response, the time dependence of the geostrophic flow has only a minor impact on the IGW scattering. Hence, the conclusions of \citetalias{kafiabad_savva_vanneste_2019} drawn by neglecting the time dependence hold for realistic slowly evolving flows. In particular, scattering by geostrophic flow does not control the frequency distribution of IGWs which, in the absence of other mechanisms, is determined by the forcing or initial conditions. This is only strictly true over a finite range of  wavenumbers $k$, since the thickness of the boundary layer increases with $k$ (see \eqref{easym}). However, at large $k$ the hypotheses of weak flow and linear waves also break down \citepalias{kafiabad_savva_vanneste_2019} and may have a larger impact than the flow time dependence (see appendix \ref{appA}  for a discussion of the restriction on $k$ imposed by the weak-flow hypothesis).  

It is worth commenting on the sharp difference between the conclusion drawn here for three-dimensional IGWs in a three-dimensional geostrophic flow and that drawn by  \citet{dong_buhler_smith_2020} in a two-dimensional set up. This difference stems from the very different geometry of the constant-frequency surfaces which are compact in dimension two (circles) and non-compact in dimension three (cones).
 In the compact case, an initial distribution of action quickly relaxes to become uniform on constant-frequency circles, then slowly spreads across these circles because of the flow time dependence. The flux of action perpendicular to the constant-frequency circles is small, but it allows for the wave frequencies to change without restriction over long time scales. In contrast, for the (non-compact) cones of the three-dimensional case, there is a non-zero action flux along cones, even in the absence of flow time dependence, corresponding to a forward cascade towards small scales.  The flux across cones introduced by the slow time dependence of the geostrophic flow acts therefore only as a small perturbation which barely affects the (non-equilibrium) stationary spectrum at finite distances along the cones.

\backsection[Funding]{We thank Oliver B\"uhler and the anonymous referees for their valuable comments. MC was supported by the MAC-MIGS Centre for Doctoral Training under grant EP/S023291/1 of the UK Engineering \& Physical Sciences Research Council (EPSRC). HAK and JV were supported by  EPSRC, grant EP/W007436/1. JV was also supported by the UK Natural Environment Research Council, grant NE/W002876/1.}

\backsection[Declaration of interests]{The authors report no conflict of interest.}

\backsection[Data availability statement]{The data that support the findings of this study are openly available in the Geophysical Fluid Dynamics collection of Edinburgh DataShare at https://doi.org/10.7488/ds/3490.}

\backsection[Author ORCID]{M. R. Cox, https://orcid.org/0000-0002-9329-3644, H. A. Kafiabad, https://orcid.org/
0000-0002-8791-9217; J. Vanneste, https://orcid.org/0000-0002-0319-589X}

\appendix
\section{Limitation of the diffusion approximation}\label{appA}
\begin{figure}
    \centering
    \includegraphics[trim = {.7cm 0cm -.7cm 0cm}]{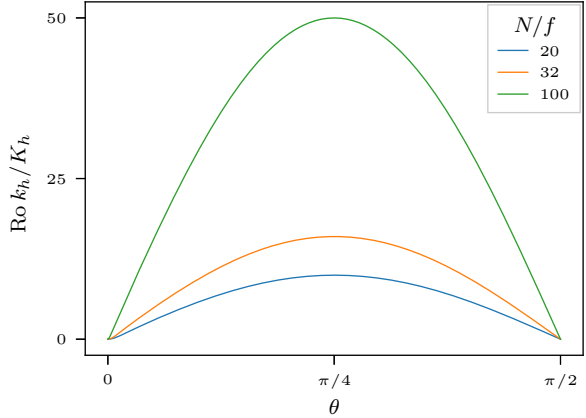}
    \caption{Upper bound of $\mathrm{Ro}\,  k_h/K_h$ given by the right-hand side of \eqref{khoverK} as a function of $\theta$ for a range of $N/f$ values including our simulation value, $N/f = 32$.} 
    \label{fig:appendix1}
\end{figure}

The diffusion approximation \eqref{diffusioneq} on which  \eqref{e_prescaling} and  \citetalias{kafiabad_savva_vanneste_2019} rely is valid for $U \ll c$. 
Defining the velocity-based Rossby number,
$\mathrm{Ro} = U K_h/f$
(rather than the vorticity-based definition of \S\ref{comparison}), and using $c$  in \eqref{groupvel},  we can rewrite this condition as
\begin{align}\label{khoverK}
    \frac{k_h}{K_h} \ll \frac{\left((N/f)^2 - 1\right)\sin^2 \theta \cos \theta}{\mathrm{Ro} \left((N/f)^2 \sin^2 \theta + \cos^2 \theta\right)^{1/2}}
\end{align}
where $k_h = k \sin \theta$ is  the horizontal wavenumber and we have taken  $0 \leq \theta \leq \pi/2$ without loss of generality. Figure \ref{fig:appendix1} displays the right-hand side of \eqref{khoverK} against $\theta$ for a range of values of $N/f$ typical of the ocean and atmosphere. The figure shows that, for realistic, small Rossby numbers ($\mathrm{Ro} \in [10^{-2},10^{-1}]$), the range of $k_h$ over which the diffusion approximation is valid extends to 20--200 times the typical flow wavenumber $K_h$ for all IGWs except those with frequencies very close to $f$ ($\theta=0$) and $N$ ($\theta=\pi/2$). 
(A scattering theory tailored to  IGWs with frequencies close to $f$, that is,  near-inertial waves, is developed in \citet{dani-v16}.)

\begin{figure}
    \centering
    \includegraphics[]{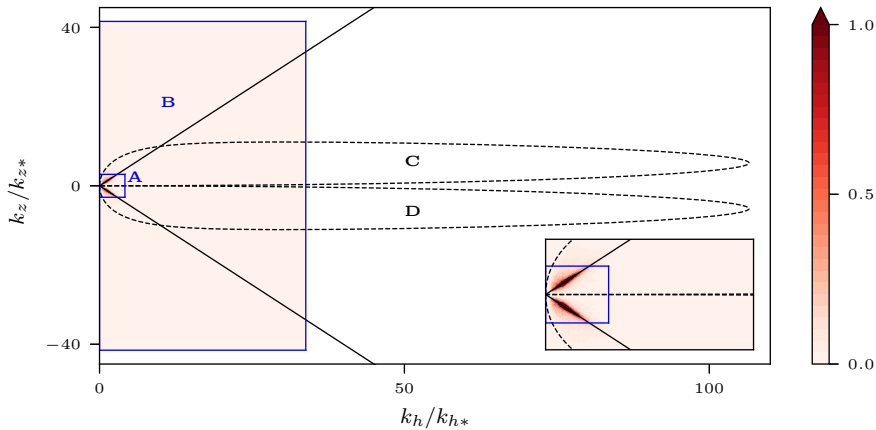}
    \caption{Region of validity of the diffusion approximation in the simulation: inequality \eqref{koverkstar} holds in regions C and D, where the diffusion approximation applies; rectangle A (also in the inset) reproduces  figure \ref{fig:contourdata} in showing the energy density $e$; rectangle B shows the ranges of $k_h$ and $k_z$ resolved in the simulation.} 
    \label{fig:appendix2}
\end{figure}

To determine the range of $k_h$ and $k_z$ for which condition \eqref{khoverK} is met in our simulation,
we recast \eqref{khoverK}  in terms of $k/k_*$ as used in  figures \ref{fig:fit} and \ref{fig:loglog}  to obtain 
\begin{align}\label{koverkstar}
    \frac{k}{k_*} = \frac{k_h \sin \theta_*}{k_{h*} \sin \theta} 
    = \frac{k_z \cos \theta_*}{k_{z*} \cos \theta} \ll  \frac{\left((N/f)^2 - 1\right)\sin \theta \cos \theta}{\mathrm{Ro} \left((N/f)^2 \sin^2 \theta + \cos^2 \theta\right)^{1/2}} \frac{K_h}{k_*}.
\end{align}
The simulation parameters are: $N/f = 32$, $k_* = 221.3$ and $K_h = 4$. The velocity-based  Rossby number is estimated to be $\mathrm{Ro} = 0.05$. Using these parameters  we compute the curve in the 
$(k_h,k_z)$-plane where  
\eqref{koverkstar} is satisfied as an equality and show the result in  figure \ref{fig:appendix2}.
The two lobes labelled C and D indicate the region of validity of the diffusion approximation.  The rectangles labelled A (also shown in the inset) and B  show the ranges of $k_h$ and $k_z$ used in figure \ref{fig:contourdata} and resolved in the simulation, respectively. This confirms that the diffusion approximation applies to the typical wavenumbers considered in our analysis.
However, because of the rapid change of $k_z$ as $\theta$ decreases from $\theta_*$, the diffusion approximation can be expected to break down around $\theta - \theta_* \approx -0.03$ in figure \ref{fig:fit}. This likely explains the mismatch between theoretical prediction and simulation results to the left of the curves' maxima in the figure.

\bibliographystyle{jfm}
\bibliography{jfm}



\end{document}